\documentclass[aip,amsmath, amssymb, reprint]{revtex4-1}
\usepackage{graphicx}
\usepackage{xcolor}
\usepackage{amsmath,amssymb}
\usepackage{float}
\usepackage{physics}
\usepackage{amsfonts}
\usepackage{verbatim}
\usepackage{balance}
\usepackage{endnotes}
\usepackage{footnote}
\usepackage{adjustbox}
\usepackage{gensymb}
\usepackage{subfigure}
\usepackage{float}
\usepackage{epigraph}
\usepackage{xcolor}
\usepackage[utf8]{inputenc}
\usepackage{setspace}
\usepackage{lipsum}
\usepackage{mwe}

\newcommand{\beq}{\begin{eqnarray}}
\newcommand{\eeq}{\end{eqnarray}}

\usepackage{amsmath}

\usepackage[normalem]{ulem}

\begin{document}

\title{Performance of the MACE-MP-0 potential for calculating viscosity in LiF molten salt.}

\author{H. L. Devereux}
\affiliation {School of Physical and Chemical Sciences, Queen Mary University of London, Mile End Road, London, E1 4NS, UK}
\author{M. Withington}
\affiliation {School of Physical and Chemical Sciences, Queen Mary University of London, Mile End Road, London, E1 4NS, UK}
\author{C. Cockrell}
\affiliation {Nuclear Futures Institute, Bangor University, Bangor, LL57 1UT, UK}
\author{K. Trachenko}
\affiliation {School of Physical and Chemical Sciences, Queen Mary University of London, Mile End Road, London, E1 4NS, UK}
\author{A. M. Elena}
\affiliation{Scientific Computing Department, Science and Technology Facilities Council, Daresbury Laboratory, Keckwick Lane, Daresbury, WA4 4AD, UK}

\begin{abstract}
We perform molecular dynamics simulations of molten Lithium Fluoride  using the MACE-MP-0 (small) machine learnt interatomic potential and the classical Buckingham and Born-Huggins-Mayer potentials. We find that the MACE-MP-0, out-of-the-box, is able to accurately reproduce the experimental viscosity across the liquid state. Whilst the previous predicted viscosities from classical potentials are under-predicted, which has previously been attributed to a suppressed melting temperature. We find that the melting temperature simulated by MACE-MP-0, simply by heating a crystal structure, is significantly closer to the experimental melting temperature of LiF.
\end{abstract}

\maketitle
\section{Introduction}
Molten salts such as LiF have received long term and recent interest due to their use in solar cells, as electrolytes and solvents \cite{papageorgiou1996performance}, and potential utilisation as coolant fluids in nuclear reactors, along with metals \cite{Rosenthal1970,FRANDSEN2020,Allen2007,Tang2015}. Water is the usual coolant in these nuclear applications but suffers from a high-reactivity \cite{US_nuclear2011}. In addition to the use of molten salts as a coolant in the reactor, there are proposals to use the salt as a carrier for the fuel, by dissolving the fuel into the salt. This would allow for adjustments to be made without unloading the core and reduce the amount of nuclear waste generated \cite{Merle2009,Locatelli2013}. Molten salts can host pyroprocessing reactions with very high activity nucleides to increase fuel efficiency and reduce waste generation \cite{Salanne2008,Locatelli2013,Uozumi2021,Lee2011,Mitachi2022}. Furthermore, molten salts have seen use as thermal storage media to complement renewable energy sources, with research interest growing in recent years \cite{Carabello2021,Bhatnagar2022,Yang2010, bauer2021molten}.

A thorough understanding of the thermodynamic and transport properties of molten salts and their mixtures will underlie the upscaling of the above processes to industrial scales, however this understanding is hindered by a poor theoretical understanding of the liquid state \cite{myreview, mybook, granato},  and the inadequacy of classical atomistic models in generating faithful thermophysical data \cite{proctor1, chen-review}.

Viscosity is one key property governing the performance of working fluids in thermal hydraulics, which has been investigated theoretically in relation to its temperature dependent minimum that may be related to fundamental physical constants \cite{sciadv1, myreview}. Previously we investigated this minimum in LiF using classical molecular dynamics (MD) \cite{Withington2024Viscosity} where, as found elsewhere \cite{LUO2016203}, the Buckingham potential model of molten LiF \cite{Buckingham1938, TOSI196445, Sangster1976Interionic, LUO2016203} predicts the viscosity at a noticeable off-set temperature. Ciccotti \textit{et al.} reported a slightly better agreement using the Born-Huggins-Mayer (BHM) potential and by analysing the response to an explicit shearing force for one temperature point \cite{Ciccotti1976Transport} as far back as 1976.

Motivated by these applications and a desire to improve the MD picture of molten LiF we now turn to the rapidly growing field of machine-learnt interatomic potentials (MLIPs). The MACE-MP-0 (MACE) MLIP \cite{Batatia2022mace, Batatia2022Design} is one example which has seen diverse successes across atomistic material modelling \cite{batatia2023foundation}. Here we report the accuracy of viscosity values calculated using the MACE potential against experimental viscosity data.
\section{Methods}
We perform MD simulations of molten LiF using classical MD (using the Buckingham and BHM potentials), \textit{via} the DL\_POLY package \cite{dlpoly-ref}, and MACE MLIP \cite{Batatia2022mace, Batatia2022Design}, \textit{via} the python package janus-core \cite{elliott_2024_14001356} using MACE\_MP\_0 (small flavour). In both cases we use the Green-Kubo method \cite{Zwanzig1965, tuckerman2023statistical, zhang2015reliable} to calculate viscosity from the integral
\begin{equation}
    \eta=\frac{V}{k_{\mathrm{B}}T}\int_0^\infty \dd t \langle \sigma_{xy}(t) \ \sigma_{xy}(0) \rangle, \label{eq:visc}
\end{equation}
where $\langle \cdot \rangle$ is the ensemble average, $T$ is the mean system temperature and $k_\mathrm{B}$ is Boltzmann's constant. Equation \ref{eq:visc} may be calculated using DL\_POLY's on-the-fly correlator. The stress tensor is defined as
\begin{equation}
    \sigma_{\alpha\beta} = \frac{1}{V}\sum_{i}m^{i}v^{i}_{\alpha}v^{i}_{\beta} - \frac{1}{V}\sum_{i, j\neq i}r_{\alpha}^{ij}F_{\beta}^{ij}, \label{eq:stress}
\end{equation}
where $\alpha$ and $\beta$ are Cartesian component indices, $m^{i}$, $v^{i}$, $r^{i}$ are particle $i$'s mass, velocity, and position, and $\mathbf{r}^{ij} = \mathbf{r}^{i} - \mathbf{r}^{j}$, $\mathbf{F}^{ij}$ is the force on $i$ due to $j$ and $V$ is the system volume. Finally, $x$ and $y$ are orthogonal Cartesian coordinates. We also compute the average viscosity utilising the other off-diagonal terms, $z-x$ and $y-z$, analogously to Equation \ref{eq:visc}. We also use the python package ASE \cite{ase-paper} to calculate the partial radial distribution functions (RDFs), $g(r)$. Then to calculate the structure factor we use \cite{Balucani1994}
\begin{equation}
    S(k) = 1 + 4\pi n \int_0^{\infty} dr\;r^2 [ g(r) - 1]\frac{\text{sin}\;kr}{kr},
\end{equation}
with $n$ the number density.

For accurate viscosity statistics a long simulation run-time is required as well as a sufficient number of independent samples \cite{zhang2015reliable, LUO2016203, Cockrell2021Universal, Withington2024Viscosity}. For all statistics collection simulations we use the NVE ensemble with a timestep of $1$ fs and a trajectory length of 1 ns, and average across $20$ initial configurations each independently initialised with the same density. For equilibration we heat the system in the NPT ensemble from a crystal lattice structure taking temperature steps from $10$ K to the desired temperature in steps of $50$ K. Each subsequent heating step lasts for 10 ps, except when determining the melting point, in proximity to which we simulate for 50 ps. We perform a final 10 ps equilibration at each temperature with 20 different initial velocity seeds in the NVE ensemble before production runs.

The viscosities calculated from Eq. \ref{eq:visc} are compared to experimental values of viscosity \cite{janz1977physical, ABE1981173, Ejima1987ViscosityOM, ana-maria}. For both MD methods we use the same maximum correlation time for calculating $\eta$, $5$ ps. Janus-core is able to make use of MACE GPU hardware acceleration for calculations, however the added computational complexity of the many-bodied MACE potential stills leads to a far longer simulation wall-time (typically one order of magnitude) than the CPU based DL\_POLY calculations based on classical potentials. In both we use a system size of $N=512$ atoms, where we have previously found little impact on results from utilising large system sizes up to 100000 atoms for classical potentials in LiF \cite{Withington2024Viscosity} and Argon \cite{Cockrell2021Universal} systems.

For the Buckingham and BHM potentials we use standard parameters for LiF stemming from the methods of Tosi and Fumi for the alkali halides of NaCl-type \cite{TOSI196445, Sangster1976Interionic, Ciccotti1976Transport, LUO2016203}. Ignoring electrostatics, in DL\_POLY the Buckingham and BHM potentials have the form
\begin{align}
    U_\mathrm{Buckingham}(r) &= A e^{-\frac{r}{\rho}}-\frac{C}{r^6}\label{buck},\\
    U_\mathrm{BHM}(r) &= Ae^{B(\sigma-r)}-\frac{C}{r^6}-\frac{D}{r^8}\label{bhm},
\end{align}
for parameters $A, B, \sigma, \rho, C$, and $D$ and inter-atomic distance $r$. The parameters for all of the interactions are given in table \ref{tab:buck}. When performing our simulations in DL\_POLY we also use the Smooth-particle Mesh Ewald method for electrostatic interactions \cite{essmann1995smooth}.
\begin{table}
    \centering
    \begin{tabular}{|l|c|c|c|}
    \hline
        Buckingham & Li-Li & Li-F & F-F \\\hline
        $A$ [eV] & 98.92 & 228.99 & 420.48 \\
        $\rho$ [\AA] & 0.299 & 0.299 & 0.299 \\
        $C$ [\AA$^{6}$ eV] &  0.046 & 0.499 & 9.051 \\\hline
        BHM & & & \\\hline
        $A$ [eV] & 0.422 & 0.290 & 0.158 \\
        $B$ [\AA$^{-1}$] & 3.344 & 3.344 & 3.344 \\
        $\sigma$ [\AA] & 1.632 & 1.995 & 2.358 \\
        $C$ [\AA$^{6}$ eV] & 0.0456 & 0.499 & 9.050 \\
        $D$ [\AA$^{8}$ eV] & 0.019 & 0.374 & 10.610 \\\hline
    \end{tabular}
    \caption{Buckingham and BHM potential parameters (in DL\_POLY format) corresponding to Equations \ref{buck} and \ref{bhm} to three decimal places.}
    \label{tab:buck}
\end{table}

The MACE model relies upon a Message Passing Neural-Network (MPNN) architecture where the input space is the dynamic local environment around any given atom along with its static properties (e.g. chemical element). Messages are constructed with radial basis functions and spherical harmonics to encode known symmetries in physical applications. Importantly the MACE model utilises higher-order many-body ACE expansion \cite{bernstein2024gap} for message passing to achieve faster and more accurate convergence to ground-truth Density Functional Theory (DFT) energies, forces, and stresses than other similar schemes to build MLIPs. The model is trained upon the MPtrj data set introduced for CHGNet \cite{deng2023chgnet} and compiled from the Material Project dataset \cite{jain2013commentary}, which includes 89 elements (including Li and F) in $\sim 146,000$ materials. Not all the MLIPs are suitable for molecular dynamics studies since not all possess conservative forces and stresses. MACE, however, is a conservative model \cite{matbench}.
\section{Results and Discussion}
\begin{figure}
    \centering    \includegraphics[width=\linewidth]{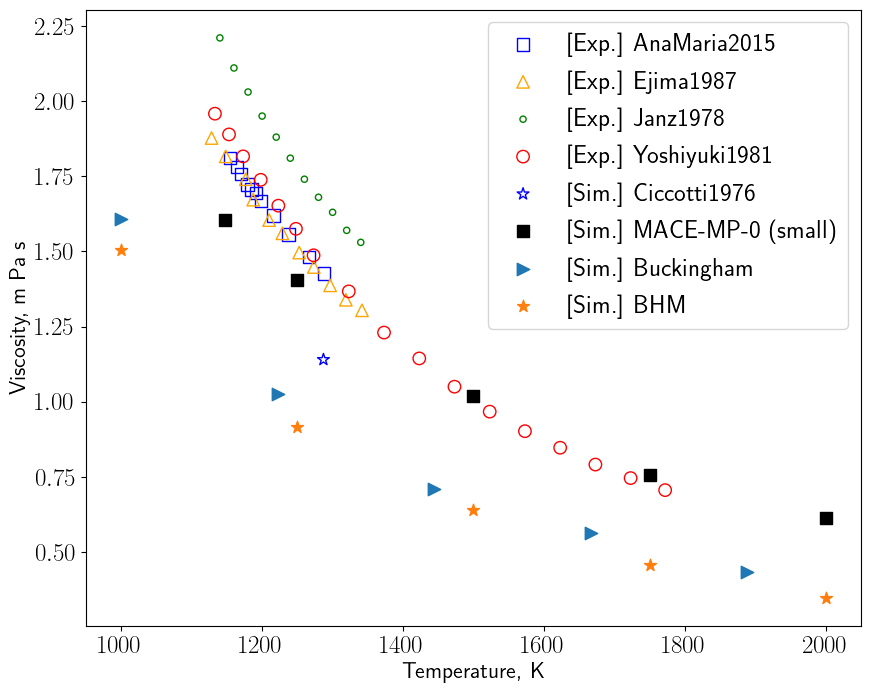}
    \caption{Viscosity, $\eta$, from experimental results \cite{janz1977physical, ABE1981173, Ejima1987ViscosityOM, ana-maria} and calculated results using the MACE, BHM \cite{Ciccotti1976Transport} and Buckingham potential. As reported previously for LiF \cite{LUO2016203, Withington2024Viscosity}, the Buckingham potential has the correct trend but is off-set on the temperature axis. The MACE results show good agreement with experimental results.}
    \label{fig:viscosity}
\end{figure}

We begin by comparing the viscosity measurements from the three computational models to experimental data in Figure \ref{fig:viscosity}. The results for the MACE model agree well with the experimental data across the temperature axis, unlike the known offset for the Buckingham and BHM models: the simulated viscosity matches the experimental viscosity well, albeit with the latter at a temperature around 300 K higher.

\begin{figure}
    \centering
    \includegraphics[width=\linewidth]{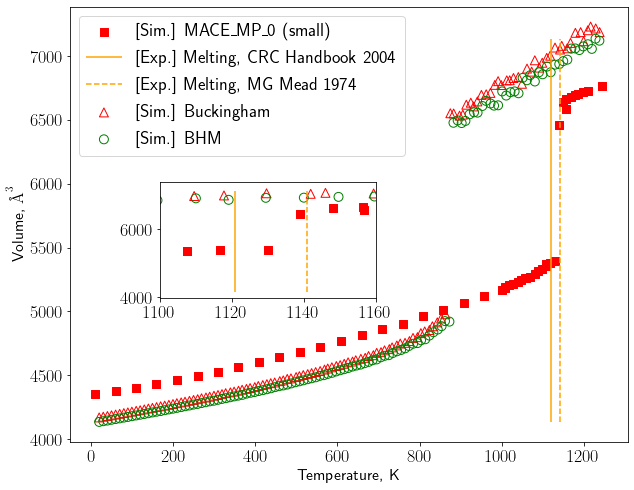}
    \caption{Volume from heating LiF from a crystal structure from $10$ K up to $1250$ K, with inset showing the melting point in more detail. The melting point is more accurately determined by the MACE potential than the Buckingham potential. CF the off setting of the viscosity data along the temperature axis in Figure \ref{fig:viscosity}. The offset of the melting point in temperature is approximately the same (200-300 K).}
    \label{fig:melting}
\end{figure}

Previously, the offsetting of viscosity values in the Buckingham results have been ascribed to the model undergoing melting at a suppressed temperature value commensurate with the gap between the simulated and experimental viscosity data \cite{LUO2016203}. To investigate this, in Figure \ref{fig:melting} we show the system volume $V$ for the MACE, Buckingham, and BHM models following a gradual monotonic heating from a crystal structure up to $1250$ K. We observe good agreement between the experimental melting temperature and the melting temperature of the MACE simulation. This supports this notion that the poor reproduction of the experimental viscosity in simulations governed by the Buckingham and BHM models is related to the underestimation of the the melting temperature by approximately $300$ K, ascertained in experiments to be $1121$ K \cite{douglas1954lithium,lide2004crc} (although 1143 K is also recorded \cite{mead1974comparison}).

It should be noted that there are several factors contributing to the location of melting point in molecular simulations. For example, simulated systems often employ periodic boundary conditions and hence have no free surfaces or interfaces, present experimentally, which promote melting. Another factor is the small system sizes in MD simulations compared to experiments. This leads to the absence of long-wavelength and low-frequency phonons with large displacement amplitudes which destabilise the solid structure and likewise encourage melting. For these and other reasons, gradual heating from a crystal structure might not be generally accurate for determining the melting point \cite{corradini2014effect, lanning2004solid,frenkel2002understanding} and other methods such as calculating the free energies of solid and liquid phases, simulating phase co-existence and so on \cite{alfe2002complementary} can be used instead. The performance of MACE potentials for determining the melting point can be tested using these approaches, however the goal and emphasis of this work is the calculation and prediction of transport properties such as viscosity rather than determining the melting point accurately which is a task for future investigations. Furthermore, the very large underestimation of the melting line by the classical potentials means the crude method we employ here is informative in gauging the improvement of the MLIP. It is nevertheless interesting to observe the good agreement between modelling and experimental melting point in Figure \ref{fig:melting}.

\begin{figure}
    \centering
    \includegraphics[width=\linewidth]{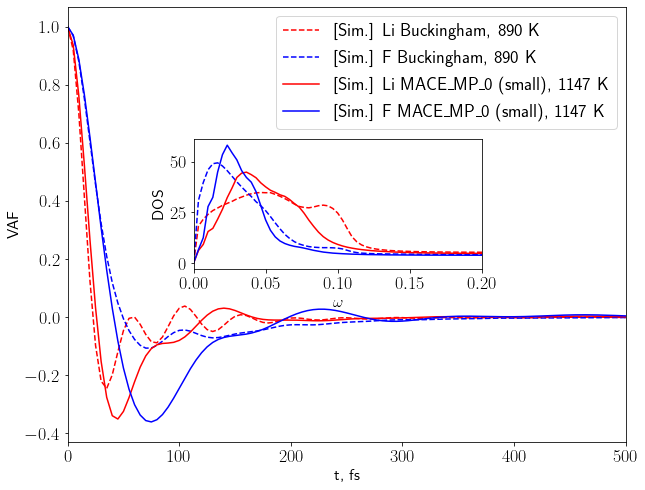}

    \caption{(a) VAFs for the Buckingham potential and MACE, along with the respective Density of States (DOS). The temperatures of 890 K and 1147 K were chosen to compare the VAFs just above the melting points of the Buckingham and MACE potentials respectively. The DOS were calculated using lag times up to $1,500$ fs.}
    \label{fig:vaf}
\end{figure}

Given the accuracy of the viscosity results of the MACE model we also examine the resulting structure and dynamics, to determine if they are physically reasonable. First we show the velocity autocorrelation functions (VAFs) of the Buckingham and MACE models in figure \ref{fig:vaf}, with the resultant densities of states (DOS) of each included as insets. We calculate these values just above the melting points in each case to compare the dynamics at the lowest temperature still consistent with the liquid state. Our DL\_POLY results for LiF are consistent with previous MD results \cite{Ribeiro2003Chemla}. The results for the MACE potential are significantly different from those of the Buckingham model. We see a decrease of the well known oscillatory frequency of F and particularly Li atoms in the MACE simulation. Both Li and F atoms exhibit a much deeper minima in the MACE VAFs, approximately equal in magnitude. Furthermore the shapes of their respective VAFs are very similar, whereas there are clear qualitative differences in the classical VAFs. The VAF minima and their depth are associated with the solidlike oscillatory component of liquid dynamics, whereas the absence of the minima signals the disappearance of the oscillatory component and purely gaslike dynamics \cite{flreview}. In this sense, deeper minima in MACE systems imply more pronounced solidlike component of the liquid motion.

\begin{figure}
    \centering
    \includegraphics[width=\linewidth]{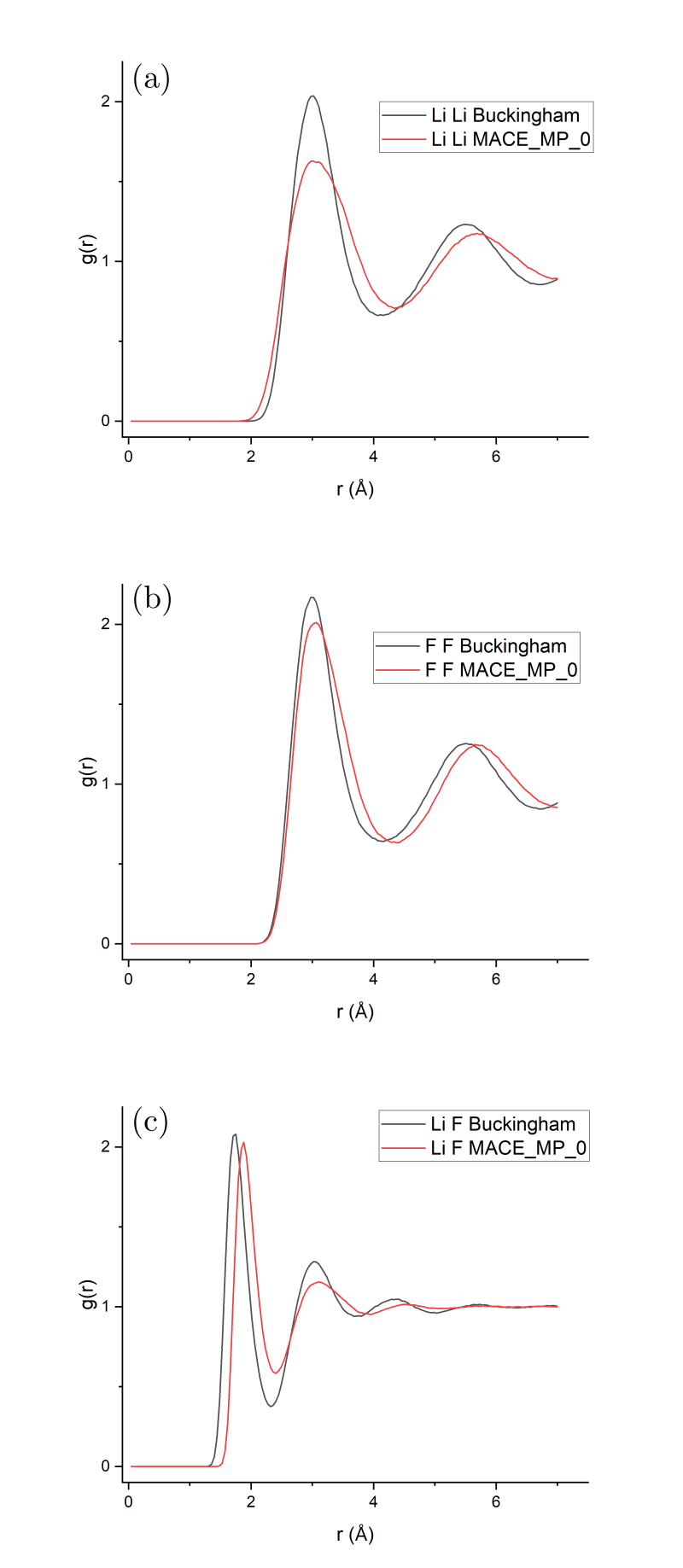}
    \caption{Partial radial distribution functions (RDF) calculated for the respective melting temperatures of the potentials. 890K for Buckingham and 1147K for the MACE.  (a) Lithium - Lithium , (b) Fluorine - Fluorine , (c) Lithium - Fluorine}
    \label{fig:rdf}
\end{figure}

\begin{figure}
    \centering
    \includegraphics[width=\linewidth]{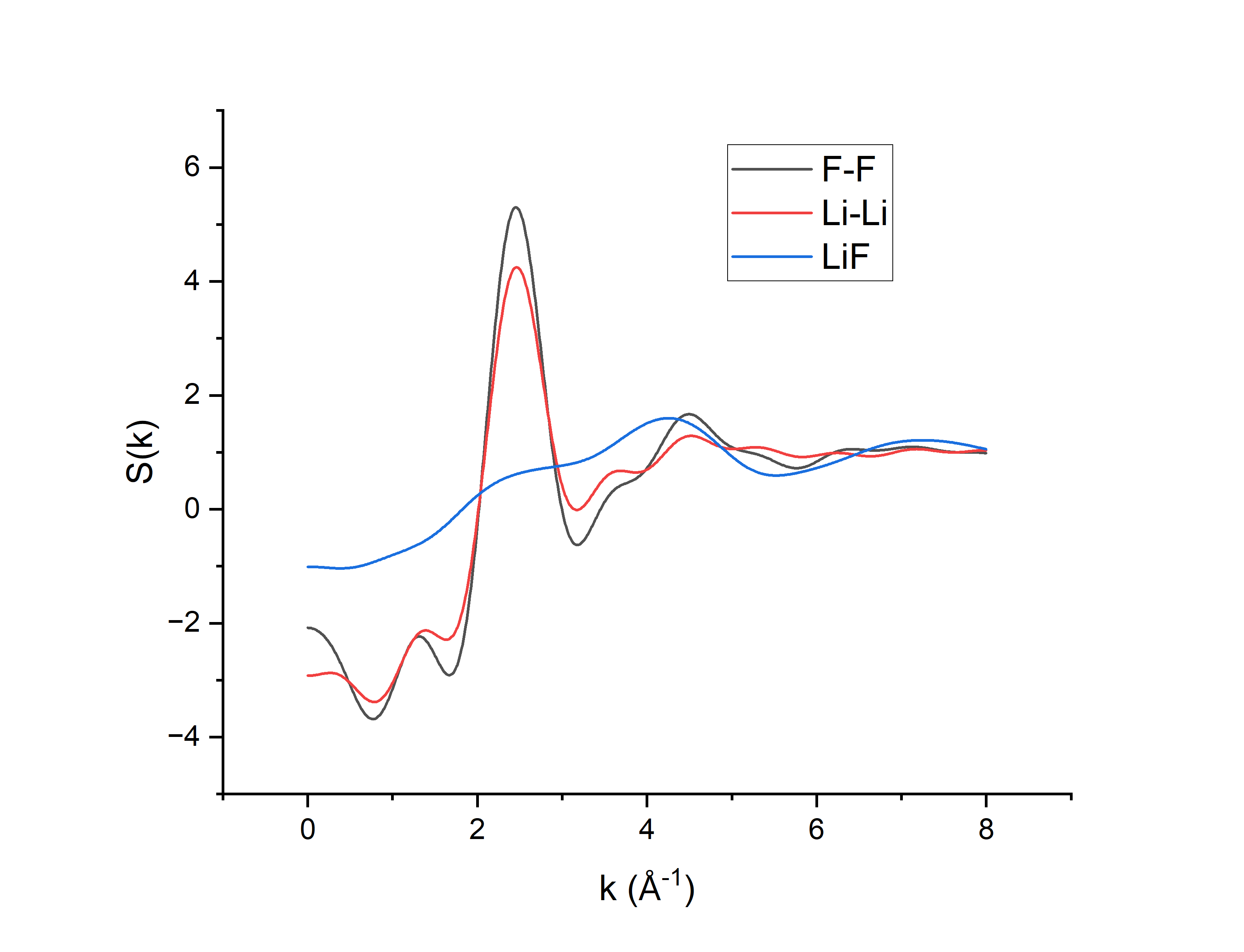}
    \caption{S(k) calculated at from the partial RDFs using 1000 MACE trajectory frames at 1147 K. }
    \label{fig:sk}
\end{figure}

Next we look at the partial RDFs $g(r)$ in Figure \ref{fig:rdf}, and the structure factors derived from them in Figure \ref{fig:sk}. We show $g(r)$ for each pair at the approximate melting temperature of each potential. The first and second peaks of each model are located at approximately the same radial separation, with the MACE peak heights slightly lower and the peak positions slightly higher than those of the classical RDFs. The reduction in magnitude indicates a somewhat broader coordination environment simulated by the MACE model than the classical model, which may be related to the lower frequency oscillation of the Li ions in the MACE simulation.
In Figure \ref{fig:rdf} we can clearly see the first two peaks for each of the partial RDFs. In order to increase the calculation cutoff radius and see the peaks reduce further and the g(r) tend to 1 we would need a larger system. For the computational time and power we have available this is not practical with MACE-MP-0. Although limited by the cut-off, the overall shape and location of peaks of $S(k)$ and $g(r)$ is consistent with the expected results for LiF simulated with MD \cite{Ribeiro2003Chemla, abramo2020structure}.

An interesting insight from the RDF analysis is that average structures can be quite similar, yet dynamics and transport properties can differ substantially as we saw for viscosity results earlier. This underscores the point which is perhaps not unexpected: similar equilibrium separations may or may not involve similar stiffness of the interatomic potentials which set the local activation energies and transport properties.

\section{Conclusions}
We have investigated the performance of the MACE MLIP applied to the MD modelling of the molten salt LiF. The MACE potential is able to more accurately reproduce experimental properties of molten LiF, but at the cost of an increased computational load versus classical two-body potentials. This is weighted against its out-of-the-box ability to obtain these results, with better results obtainable from a dedicated potential with more training data specific to the molten phase of LiF.

Our main interest was in the prediction of viscosity, using the same method as has been used in extant MD potentials (Buckingham and BHM) that is known to produce the correct trend but offset along the temperature axis for LiF. We have found that the MACE potential agrees well with experimental viscosities over a range of temperatures above melting. Further we have found that, when applying a na\"ive monotonic heating methodology to crystalline LiF, the MACE potential is also able to recover the experimental melting temperature surprisingly accurately given the difficulty of ascertaining melting points in MD and DFT. The large discrepancy between the melting points of classical potentials and experiments has previously been the given explanation for the offset viscosity values. Our results lend some credence to that hypothesis, but it remains possible other effects such as system size and the lack of interfaces may contribute.

Looking to the system's structure and dynamics, our results indicate that the MACE potential generates a broader coordination environment around each ion, possibly reducing the stiffness of the aggregate interactions and thereby causing the decreased frequency in the oscillation of the Li atoms. The dynamics of the anions and cations are more similar in the MACE simulations, while in classical MD simulations they are quite distinct.

The far superior prediction of the viscosity and melting temperature of LiF indicate that the dynamics resulting the MACE potential, which differ qualitatively from those of classical potentials, are indispensable to the accurate simulation of this system. Dedicated potentials for the simulation of other molten salts and their mixtures will help guide experimentation and inform their application to industry.

\section*{Acknowledgements}
We are grateful to EPSRC (for grant No. EP/W029006/1).
This research utilised Queen Mary's Apocrita HPC facility, supported by QMUL Research-IT http://doi.org/10.5281/zenodo.438045, STFC Scientific Computing Department’s SCARF cluster and the Sulis Tier 2 HPC platform hosted by the Scientific Computing Research Technology Platform at the University of Warwick and funded by EPSRC GrantS EP/T022108/1 and EP/V028537/1 and the HPC Midlands+ consortium.

\bibliographystyle{unsrtnat}
\bibliography{bib.bib}
\end{document}